\documentclass[pra,twocolumn,tightenlines,showpacs,nofootinbib]{revtex4}
\usepackage{multirow}
\usepackage{amsmath}
\usepackage{bm}
\usepackage{dcolumn}

\begin{document}
\title{Study of quadrupole polarizabilities with combined configuration interaction
       and coupled-cluster method}
\author{S.~G.~Porsev$^{1,2}$}
\author{M.~S.~Safronova$^1$}
\author{M.~G.~Kozlov$^2$}
\affiliation{
$^1$Department of Physics and Astronomy, University of Delaware,
    Newark, Delaware 19716, USA\\
$^2$Petersburg Nuclear Physics Institute, Gatchina,
    Leningrad District, 188300, Russia}

\date{\today}

\begin{abstract}
The recently developed method combining the configuration interaction and the coupled-cluster method was demonstrated to provide accurate treatment of correlation corrections in divalent atomic systems [M.~S.~Safronova, M.~G.~Kozlov, and C.~W.~Clark, Phys. Rev. Lett. {\bf 107}, 143006 (2011)]. We have extended this approach to the calculation of quadrupole polarizabilities $\alpha_2$ and applied it to evaluate $\alpha_2$ for the ground state of  Mg and Mg-like Si$^{2+}$. Performing the calculations in three different approximations of increasing accuracy allowed us to place the upper bounds on the uncertainty of the final results. The recommended values $\alpha_2(3s^2\,^1\!S_0)= 35.86(13)$ a.u. for Si$^{2+}$ and $\alpha_2(3s^2\,^1\!S_0)= 814(3)$ a.u. for Mg are estimated to be accurate to 0.37\%. Differences in quadrupole polarizability contributions in neutral Mg and Si$^{2+}$ ion are discussed.
\end{abstract}
\pacs{31.15.ap, 31.15.ac, 31.15.am, 31.15.bw}
\maketitle

\section{Introduction}

Atomic polarizabilities are important for a number of applications ranging from studies of fundamental symmetries to development of ultra-precise
atomic clocks as well as modeling of properties of chemical compounds. As a result, the study of polarizabilities has been of increasing importance
in recent years. A number of both experimental and theoretical methods exist for determination of E1 electric-dipole polarizabilities. We refer the
reader to a recent review \cite{MitSafCla10} and references therein for the discussion of the applications of E1 polarizabilities and methodologies
to determine these quantities in systems with a few valence electrons.

The quadrupole polarizabilities have been a subject of far lesser number of studies, and few high-precision values exist. The previous theoretical
studies used the exponentially correlated Gaussian functions~\cite{Kom02}, the pseudopotential methods \cite{PatTan98,MitBro03,MitSaf09},
coupled-cluster methods~\cite{SafSaf08,MitSaf09,Sah07}, and combined the configuration interaction and the many body perturbation theory (CI+MBPT)
\cite{PorDer06Z}. Most of these studies have been limited to either very light or monovalent atomic systems. There are ever fewer experimental studies of the quadrupole polarizabilities  owing to intrinsic difficulty in measuring this property.

 The quadrupole polarizability term arises in the effective
potential for polarization interactions between the core and the Rydberg electrons which allows determination of $\alpha_2$ from the analysis of the
fine-structure intervals of the high-L Rydberg states using a resonant excitation Stark ionization spectroscopy (RESIS) method \cite{Lun05}. This
approach has been used to determine the ground state quadrupole polarizabilities of Mg$^{+}$~\cite{SnoLun08}, Ba$^+$~\cite{SnoGeaKom05,SnoLun07}, and
Th$^{4+}$~\cite{HanKeeLun10,KeeLunFeh11}. The lowest electric-quadrupole matrix elements involving the ground state, which generally give the
dominant contribution to $\alpha_2$ have been determined using the RESIS method for the $6s-5d_{j}$ transitions in Ba$^+$~\cite{WooLunSno09} and
$5f_{5/2}-6d_{j}$ transitions in Th$^{3+}$ ~\cite{KeeHanWoo11}.

The difficulty of determining the quadrupole polarizability from RESIS experiments that led to initial disagreement with theory in Ba$^+$ was analyzed in \cite{SnoLun07}. The discrepancy between theory and experimental values was resolved in the same work \cite{SnoLun07}. The Mg$^+$  experimental $\alpha_2$ value from \cite{SnoLun08} was found to be in disagreement with both coupled-cluster and configuration interaction with semi-empirical
core polarization potential (CICP) results ~\cite{MitSaf09}, while both theoretical values are in excellent agreement with each other~\cite{MitSaf09}. The revised polarization plot analysis of Ref.~\cite{SnoLun08} data carried out in \cite{MitSaf09} yielded much lower value, that is only slightly outside of the combined error estimates. The RESIS Th$^{4+}$ $\alpha_2$ value is in agreement with theory predictions taking into account the
uncertainties, but the theoretical calculations are of low accuracy for this ion.

 In summary, RESIS method can be used to determine the quadrupole
polarizabilities to a good precision, but more benchmark cases of comparison with high-precision theory are needed. Therefore, it is important to
develop approaches that are able to predict quadrupole polarizabilities to high precision and evaluate the uncertainties of the final values. We note
that quadrupole polarizability is a particularly good property for benchmark testing of theoretical approaches owing to generally large contributions
of the correlation corrections. Accurate calculations and benchmark tests of divalent atomic system properties are of particular interest owing to
applications to atomic clock research~\cite{RosHumSch08}, fundamental symmetry studies~\cite{TsiDouFam09}, and quantum
information~\cite{GorReyDal09}.

The method combining the configuration interaction (CI) and the all-order coupled-cluster method (CI+all-order) developed  in
Refs.~\cite{Koz04,SafKozJoh09}  was applied to the calculation of the electric-dipole static polarizabilities and the corresponding blackbody
radiation shifts in divalent B$^+$, Al$^+$, In$^+$, and Tl$^+$ in~\cite{SafKozCla11,ZurSafKoz12}. We applied the same approach to calculate
electric-dipole polarizabilities of the several low-lying states of Si$^{2+}$ ion in Ref.~\cite{SafPorKoz12}. These polarizabilities were calculated
with unprecedented  $\sim 0.1\%$  accuracies demonstrating a great potential and a high efficiency of the CI + all-order method in calculation of
such quantities. Therefore, we select the ground state electric-quadrupole (E2) polarizabilities of Mg and Si$^{2+}$ as the benchmark test cases for
the approach developed in the present work, since the high precision of the theory is expected for these cases. We note that the method developed in
this work can be used to calculate quadrupole polarizabilities of the low-lying states of any divalent or trivalent system.

The E2 polarizability of the Mg ground state was calculated  using the CI+MBPT method in Ref.~\cite{PorDer06Z}. The core-valence correlations were
included explicitly in the second order of MBPT, while the higher orders of MBPT were included using the energy fitting described in
Ref.~\cite{PorDer06Z}. The final result was $\alpha_2(3s^2 ~^1\!S_0) = 812(6)$ a.u.. In this work, we carried out {\it pure} ab initio calculation of
this quantity in the framework of the CI + all-order approach. The resulting \textit{ab initio} value 814(3) a.u. is in an excellent agreement with
the result reported earlier. A small difference between the central values can be attributed to the Breit interaction included in the present work,
\textit{ab initio} treatment of the higher-order correlation corrections,
the greater basis set, and the greater CI space. For Si$^{2+}$, we obtained $\alpha_2(3s^2 ~^1\!S_0) = 35.86(13)$ a.u. We will discuss the details of the method
and evaluation of uncertainties in the following sections.

To the best of our knowledge, there are no experimental data for the  $\alpha_2(3s^2~ ^1\!S_0)$ for Si$^{2+}$ and Mg. At the same time the
theoretical accuracy of calculation of these quantities achieved in this work is sufficiently high and, respectively, the E2 polarizabilities of the
($3s^2\, ^1\!S_0$) state in Si$^{2+}$ and Mg present a good opportunity for a high-precision benchmark test of theory and experiment. Since Si$^{2+}$
electric-dipole polarizability was already accurately determined using the RESIS experiment~\cite{SnoLun07}, it is a likely candidate for a
benchmark test of determination of E2 polarizability by the RESIS method. Mg is also of particular interest since it is considered to be a good candidate for the development of the atomic clock.

\section{Method}
The details of the CI+all-order method were presented in~\cite{SafKozJoh09,SafKozCla11,SafPorKoz12}, therefore, we provide only a brief outline of
the approach. We start from the solution of the Dirac-Fock (DF) equations
$$ \hat H_0\, \psi_c = \varepsilon_c \,\psi_c, $$
where $H_0$ is the relativistic DF Hamiltonian \cite{DzuFlaKoz96b,SafKozJoh09} and $\psi_c$
and $\varepsilon_c$ are single-electron wave functions and energies.
The dominant part of the Breit interaction was included in the
self-consistency procedure~\cite{KozPorTup00}.

The calculation was carried out in V$^{N-2}$ approximation, i.e. the self-consistent calculations were performed for the [$1s^2 2s^2 2p^6$] closed
core, and  the $3s$, $3p$, $3d$, $4s$, $4p$, and $4d$ orbitals were formed in this potential. The B-spline basis set was formed in a spherical cavity
with radius 60 a.u. and consisted of $N=35$ orbitals for each partial wave up to $l=5$. The set of configurations was constructed by single and
double excitations of the electrons from the main configuration $3s^2$ to the $4s-23s$, $3p-23p$, $3d-23d$, $4f-23f$, and $5g-23g$ orbitals. Thus,
five partial waves with the orbitals having the principal quantum numbers $n\leq 23$ were involved in the construction of the CI space. We have
verified that such  CI space is numerically complete.

The wave functions and the low-lying energy levels were determined by solving the multiparticle
relativistic equation for two valence electrons~\cite{KotTup87}:
\begin{equation}
H_{\rm eff}(E_n) \Phi_n = E_n \Phi_n.
\label{Heff}
\end{equation}

The effective Hamiltonian was defined as $$ H_{\rm eff}(E) = H_{\rm FC} + \Sigma(E),$$ where $H_{\rm FC}$ is the Hamiltonian in the frozen-core
approximation. The energy-dependent operator $\Sigma(E)$ takes into account virtual core excitations. It is constructed using the second order many-body
perturbation theory in the CI+MBPT approach \cite{DzuFlaKoz96b} and linearized coupled cluster single-double method in the CI+all-order
approach \cite{SafKozJoh09}. The construction of the effective Hamiltonian in the CI+MBPT and CI+all-order approximations was described in detail in
Refs.~\cite{DzuFlaKoz96b,SafKozJoh09}.

The electric quadrupole polarizability $\alpha_2$ can be represented in a general case
as a sum of three parts
\begin{equation}
\alpha_2 = \alpha_2^v + \alpha_2^c + \alpha_2^{vc},
\end{equation}
where $\alpha_2^v$ includes excitations of valence electrons, $\alpha_2^c$ is the ionic core polarizability, and $\alpha_2^{vc}$ is the small
correction to $\alpha_2^c$ which subtracts out the excitations of core electrons into the occupied valence shells forbidden by the Pauli principle. In
our case, $\alpha_2^{vc}=0$ because there are no $nd$ orbitals in the core and, respectively, no excitations from the core to the $3s$ shell. The
ionic core polarizability, $\alpha_2^c$, was evaluated in both the DF approximation and the RPA approximation. The difference between two these
values can be used to estimate the uncertainty of this quantity.

The static electric-quadrupole polarizability of the $3s^2~^1\!S_0$ state
can be written as
\begin{equation}
\alpha_2 =
2 \sum_n \frac{\langle ^1\!S_0|Q_{0}|n \rangle \langle n|Q_{0}|^1\!S_0 \rangle}{E_n-E_0}.
\end{equation}
In atomic units (m=$\hbar$=$|e|$=1) the electric quadrupole moment operator is determined
as $Q_{\nu}= -r^2\,C_{2\nu}(\bf n)$, where ${\bf n} \equiv {\bf r}/r$
and $C_{2\nu}(\bf n)$ are the normalized spherical harmonics.

The valence part of the polarizability, $\alpha_2^v$, of the state  $|0\rangle$
can be found by solving the inhomogeneous
equation in the valence space, which is written as~\cite{KozPor99a}
\begin{equation}
(H_{\textrm{eff}} - E_0)|\Psi\rangle = Q_{\textrm{eff}} |0\rangle
\label{inhom}
\end{equation}
and then calculating
\begin{equation}
\alpha_2^v = 2 \, \langle 0|(Q_0)_{\textrm{eff}}|\Psi \rangle.
\label{alph2}
\end{equation}
The effective quadrupole operator $Q_{\textrm{eff}}$ includes the random-phase approximation
(RPA) corrections.

Our analysis shows that the RPA corrections to values of $\alpha_2(3s^2~^1\!S_0)$ for both Si$^{2+}$ and Mg are very small (a few tenth of a
percent). Therefore, all other smaller corrections to the effective operator including core-Brueckner, two-particle, structure radiation, and
normalization corrections can be omitted at the present level of accuracy.

We find that some caution is required in calculating the E2 polarizabilities by solving the inhomogeneous equation in the valence space. First, the
wave functions $|\Psi\rangle$ and $|0\rangle$ are of the same parity. Second, the wave function $|0\rangle$ is the solution of the homogeneous
equation
\begin{equation}
(H_{\textrm{eff}}-E_0)|0\rangle = 0.
\label{hom}
\end{equation}

It is known that a general solution of an inhomogeneous equation is a sum of a particular solution of
the inhomogeneous equation and the general solution of the homogeneous equation. In our case the solution $|\Psi\rangle$ of Eq.~(\ref{inhom}) is
the sum of the particular solution of Eq.~(\ref{inhom}) which we denote as $|\Psi'\rangle$ and a
solution of Eq.~(\ref{hom}):
\begin{equation}
|\Psi\rangle = |\Psi'\rangle + \beta\, |0\rangle,
\end{equation}
where $\beta$ is a numerical coefficient and $|\Psi'\rangle$ is assumed to be orthogonal to $|0\rangle$, i.e. $\langle 0|\Psi' \rangle =0$.

To separate out the particular solution $|\Psi'\rangle$ from the general solution, one needs to project $|\Psi\rangle$ to the subspace orthogonal to
$|0\rangle$ as
\begin{equation}
|\Psi'\rangle = |\Psi\rangle - |0\rangle \langle 0|\Psi\rangle.
\end{equation}

We emphasize that such an admixture does not occur in calculating E1 polarizabilities, parity nonconserving amplitudes and other quantities for which
$|\Psi\rangle$ and $|0\rangle$ are of opposite parity. For those operators, $|\Psi\rangle$ and $|0\rangle$ belong to different subspaces from very
beginning and automatically turn out to be orthogonal to each other (i.e., $\beta=0$).

It seems that if $|0\rangle = |^1\!S_0\rangle$ we do not need to worry about the admixture of $|0\rangle$ to $|\Psi'\rangle$ because $\langle
0|(Q_0)_{\textrm{eff}}|0 \rangle = 0$ and this admixture is removed from the final result. The problems arise  because the factor $\beta$ can be
very large. In particular, in our case $\beta$ was $\sim 10^5$ which would lead to a numerical instability of the method and, finally, to a loss of
accuracy in the straightforward implementation of the approach described above.

We find a solution to this problem that can be implemented in the framework of our approach without additional modifications of the  method. The solution of
the inhomogeneous equation, $|\Psi\rangle$, can be represented as a sum of projections to the states with definite total angular momenta and written
as~\cite{KozPor99a}
\begin{equation}
|\Psi \rangle = \sum_{J'=J_{\rm min}}^{J+2} |\Psi_{J',M}\rangle,
\label{Psi_J}
\end{equation}
where $J$ and $M$ are the total angular momentum and its projection of the state $|0\rangle$ and $J_{\rm min} \equiv {\rm max}(0,J-2)$. In our case,
the only term in Eq.~(\ref{Psi_J}) that is of interest for us is $|\Psi_{J'=2,M}\rangle$. All other terms do not contribute to Eq.~(\ref{alph2}).
Thus, if we find the solutions of Eq.~(\ref{Heff}) belonging only to the subspace $J'=2, M=2$
we avoid the problem discusses above because the $|\Psi_{J'=2,M=2}\rangle$ and $|0\rangle = |^1\!S_0\rangle$ have different total angular momenta and
cannot admix to each other.

\section{Results}
First, we find the low-lying energy levels of Mg and Si$^{2+}$. To estimate the accuracy of calculations we calculated the energy levels in the CI,
CI+MBPT, and CI+all-order approximations. The results for Si$^{2+}$ were presented in ~\cite{SafPorKoz12}, where we demonstrated  that the CI energy
levels were already in good agreement with the experimental values. The maximum difference between the CI  and experimental results did not exceed
2.2\%. For Mg, the agreement with experiment at the CI level is only slightly worse. An inclusion of the core-valence correlations in the CI+MBPT and
CI+all-order calculations led to further substantial improvement of the theoretical energy levels.

In Tables~\ref{table:Si2} and~\ref{table:Mg}, we present the results obtained in the CI+all-order approximation for Si$^{2+}$ and Mg, respectively,
and compare them with experimental data. The two-electron binding energies are given in the first row of these tables, the energies in other rows are
counted from the ground state.
\begin{table}
\caption{Comparison between experimental
\cite{RalKraRea11} and theoretical energy levels of Si$^{2+}$ in cm$^{-1}$.
Two-electron binding energies are given in the first row, energies in
other rows are counted from the ground state. Results of the CI+all-order
calculations are given in column labeled CI+All. Relative difference of this
calculation with experiment is given in the last column in \%.}
\label{table:Si2}
\begin{ruledtabular}
\begin{tabular}{lrccccccrrr}
  \multicolumn{1}{c}{State}
& \multicolumn{1}{c}{Expt.}
& \multicolumn{1}{c}{CI+All}
& \multicolumn{1}{c}{Difference (\%)} \\
\hline
$3s^2\,^1\!S_0 $    &  634232  &  634226 & $-$0.001\% \\
$3p^2\,^1\!D_2 $    &  122215  &  122294 &    0.065\% \\
$3p^2\,^3\!P_0 $    &  129708  &  129753 &    0.035\% \\
$3p^2\,^3\!P_1 $    &  129842  &  129887 &    0.035\% \\
$3p^2\,^3\!P_2 $    &  130101  &  130145 &    0.034\% \\
$3s3d\,^3\!D_3 $    &  142944  &  142944 &    0.000\% \\
$3s3d\,^3\!D_2 $    &  142946  &  142946 &    0.000\% \\
$3s3d\,^3\!D_1 $    &  142948  &  142948 &    0.000\% \\
$3s4s\,^3\!S_1 $    &  153377  &  153403 &    0.017\% \\
$3p^2\,^1\!S_0 $    &  153444  &  153613 &    0.110\% \\
$3s4s\,^1\!S_0 $    &  159070  &  159116 &    0.029\% \\
$3s3d\,^1\!D_2 $    &  165765  &  165898 &    0.080\%  \\
[0.2pc]
$3s3p\,^3\!P^o_0 $  &  52725   &  52770  &    0.086\%  \\
$3s3p\,^3\!P^o_1 $  &  52853   &  52897  &    0.083\%  \\
$3s3p\,^3\!P^o_2 $  &  53115   &  53159  &    0.082\%  \\
$3s3p\,^1\!P^o_1 $  &  82884   &  82933  &    0.058\% \\
$3s4p\,^3\!P^o_0 $  &  175230  &  175249 &    0.011\%  \\
$3s4p\,^3\!P^o_1 $  &  175263  &  175282 &    0.011\%  \\
$3s4p\,^3\!P^o_2 $  &  175336  &  175355 &    0.011\%  \\
$3s4p\,^1\!P^o_1 $  &  176487  &  176511 &    0.013\%  \\
\end{tabular}
\end{ruledtabular}
\end{table}
\begin{table}
\caption{Comparison between experimental
\cite{RalKraRea11} and theoretical energy levels of Mg in cm$^{-1}$.
Two-electron binding energies are given in the first row, energies in
other rows are counted from the ground state. Results of the CI+all-order
calculations are given in column labeled CI+All. Relative difference of this
calculation with experiment is given in the last column in \%.}
\label{table:Mg}
\begin{ruledtabular}
\begin{tabular}{lrccccccrrr}
  \multicolumn{1}{c}{State}
& \multicolumn{1}{c}{Expt.}
& \multicolumn{1}{c}{CI+All}
& \multicolumn{1}{c}{Difference (\%)} \\
\hline
$3s^2\,^1\!S_0 $    & 182939  &  188288 &  0.03\% \\
$3s4s\,^3\!S_1 $    &  41197  &   41184 &  0.03\% \\
$3s4s\,^1\!S_0 $    &  43503  &   43491 &  0.03\% \\
$3s3d\,^1\!D_2 $    &  46403  &   46388 &  0.03\% \\
$3s3d\,^3\!D_2 $    &  47957  &   47933 &  0.05\% \\
$3s3d\,^3\!D_3 $    &  47957  &   47933 &  0.05\% \\
$3s3d\,^3\!D_1 $    &  47957  &   47933 &  0.05\% \\
$3s5s\,^3\!S_1 $    &  51873  &   51854 &  0.04\% \\
$3s5s\,^1\!S_0 $    &  52556  &   52541 &  0.03\% \\
$3s4d\,^1\!D_2 $    &  52047  &   53114 &  0.04\%  \\
[0.2pc]
$3s3p\,^3\!P^o_0 $  &  21850  &  21849  &  0.01\%  \\
$3s3p\,^3\!P^o_1 $  &  21870  &  21869  &  0.01\%  \\
$3s3p\,^3\!P^o_2 $  &  21911  &  21909  &  0.01\%  \\
$3s3p\,^1\!P^o_1 $  &  35051  &  35044  &  0.02\% \\
$3s4p\,^3\!P^o_0 $  &  47841  &  47823  &  0.04\%  \\
$3s4p\,^3\!P^o_1 $  &  47844  &  47826  &  0.04\%  \\
$3s4p\,^3\!P^o_2 $  &  47851  &  47833  &  0.04\%  \\
$3s4p\,^1\!P^o_1 $  &  49347  &  49328  &  0.04\%  \\
\end{tabular}
\end{ruledtabular}
\end{table}
We  find that the agreement between theoretical and experimental energy levels listed in these tables is extremely good, 0.05\% or better for most of
the levels. This is important for calculation of the quadrupole polarizability of the ground state because the low-lying levels give main
contribution to this quantity. We note that the inclusion of the $ng$ orbitals to the CI space is essential to obtain such high accuracy for all
energy levels, including the singlet states, for Si$^{2+}$. For Mg the $ng$ orbitals can be omitted from the CI space with negligible loss of accuracy.

In Table~\ref{tabE2} we list the contributions of several low-lying states to $\alpha_2(^1S_0)$ for Si$^{2+}$ and Mg. We also present the absolute
values of the reduced matrix elements $|\langle ^1\!S_0 ||Q|| n \rangle|$. For Si$^{2+}$, the two transitions from the ground state to the $3p^2\,^1\!D_2$ and $3s3d\,^1\!D_2$ states contribute 97\% of the final value. Such a large contribution of the $3p^2\,^1\!D_2$ state appears at first to be surprising because the
$3s^2\,^1\!S_0 - 3p^2\,^1\!D_2$ is a two-electron transition. However, there is the large admixture (33\% in probability) of the $3s3d$
configuration to the $3p^2$ configuration that explains such a large contribution of the $3p^2\,^1\!D_2$ state to $\alpha_2(^1\!S_0)$.

The breakdown of the E2 contributions is different for Mg, where the low-lying $3snd\,^1\!D_2$ states with $n=3-5$, listed in Table~\ref{tabE2}, contribute 86\% to $\alpha_2(^1\!S_0)$. For neutral Mg, whose electrons are more weakly bound to the nucleus, the contribution of the high-lying
discrete states as well as the continuum is larger than for doubly-ionized Si. Therefore, using a sum-over-states approach with a few low-lying
contributions in the sum over intermediate states allows to obtain the ground state E2 polarizability for Mg only with an accuracy of about 15\% percent. To obtain this quantity with a higher accuracy a more sophisticated approach like a solution of inhomogeneous equation used in this work is required.
\begin{table}
\caption{The contributions to the $3s^2\,^1\!S_0$ E2 polarizabilities (in a.u)
in the CI + all-order approximation. The dominant contributions to
the valence polarizabilities are listed separately with the corresponding absolute values of
electric quadrupole reduced matrix elements given in columns
labeled $Q$. The theoretical and experimental NIST transition energies are given in columns
$\Delta E_{\rm th}$  and $\Delta E_{\rm exp}$. The
remaining contributions to $\alpha_2(^1S_0)$ are given in rows Other.
The contributions from the core are given in rows $\alpha_2^c$.
The dominant contributions to $\alpha_2$ are
calculated with the CI + all-order energies.}
\label{tabE2}
\begin{ruledtabular}
\begin{tabular}{clrrcr}
          & \multicolumn{1}{c}{Contribution}
                                             & \multicolumn{1}{c}{$\Delta E_{\rm exp}$}
                                                       & \multicolumn{1}{c}{$\Delta E_{\rm th}$}
                                                                & \multicolumn{1}{c}{$Q$}
                                                                        & \multicolumn{1}{c}{$\alpha_2$} \\
\hline \\
Si$^{2+}$ & $3s^2\,^1\!S_0 - 3p^2\,^1\!D_2$  & 122215  & 122276 & 5.200 & 19.41 \\
          & $3s^2\,^1\!S_0 - 3p^2\,^3\!P_2$  & 130137  & 130101 & 0.119 &  0.01 \\
          & $3s^2\,^1\!S_0 - 3s3d\,^3\!D_2$  & 143028  & 142946 & 0.002 &  0.00 \\
          & $3s^2\,^1\!S_0 - 3s3d\,^1\!D_2$  & 166247  & 165765 & 5.373 & 15.24 \\
          & Other                            &         &        &       &  1.09 \\
          & $\alpha_2^c$                     &         &        &       &  0.11 \\
          & Total                            &         &        &       & 35.86 \\
[0.3pc]
Mg        & $3s^2\,^1\!S_0 - 3s3d\,^1\!D_2$  & 46390   & 46403  & 18.65 & 658.6 \\
          & $3s^2\,^1\!S_0 - 3s3d\,^3\!D_2$  & 47939   & 47957  &  0.01 &   0.0 \\
          & $3s^2\,^1\!S_0 - 3s4d\,^1\!D_2$  & 53105   & 53135  &  5.05 &  42.2 \\
          & $3s^2\,^1\!S_0 - 3s5d\,^1\!D_2$  & 56289   & 56308  &  1.52 &   3.6 \\
          & Other                            &         &        &       & 109.0 \\
          & $\alpha_2^c$                     &         &        &       &   0.5 \\
          & Total                            &         &        &       & 813.9 \\
\end{tabular}
\end{ruledtabular}
\end{table}

The E2 polarizabilities of the ground state obtained in different
approximations are given for Si$^{2+}$ and Mg in Table~\ref{tab:alph2}.
\begin{table}
\caption{Comparison of the present recommended values of the ground state static E2 polarizabilities (in a.u.)
in Mg and Si$^{2+}$ with other calculations. First three rows give \textit{ab initio} results for the valence polarizabilities $\alpha_2^v$ calculated in the CI, CI+MBPT, and CI+all-order approximations. In the row $\Delta$(MBPT $-$ All) the differences of the CI+MBPT and CI+all-order values are presented. The contributions
from the core are given in the row $\alpha_2^c$. The ``Total'' are the values obtained as the sum of the CI+All values and $\alpha_2^c$.}
\label{tab:alph2}
\begin{ruledtabular}
\begin{tabular}{lcc}
                       & \multicolumn{1}{c}{Si$^{2+}$} & \multicolumn{1}{c}{Mg} \\
\hline
 CI                    & 37.58                         & 888.4   \\
 CI+MBPT               & 35.88                         & 817.5   \\
 CI+All                & 35.75                         & 814.3  \\
$\Delta$(MBPT $-$ All) &  0.13                         &   3.2    \\
 $\alpha_{2}^c$          &  0.11                         &   0.52 \\
 Total                 & 35.86                         & 813.9  \\
 Recommended value     & 35.86(13)                     & 814(3) \\
[0.1pc]
 Other works           & 35.74(36)\footnotemark[1]     & 812(6)\footnotemark[2]\\
                       &                               & 813.9\footnotemark[3]\\
                       &                               & 809.3\footnotemark[4]\\
                       &                               & 828\footnotemark[5]\\
\end{tabular}
\end{ruledtabular}
\footnotemark[1]{Ref.~\cite{Mit08}};
\footnotemark[2]{Ref.~\cite{PorDer06Z}};
\footnotemark[3]{Ref.~\cite{MitBro03}};
\footnotemark[4]{Ref.~\cite{ArcTha91}};
\footnotemark[5]{Ref.~\cite{MaeKut79}}.
\end{table}
This table illustrates that the core-valence correlations included in the second order of MBPT in the CI+MBPT approximation and in all orders in the
CI+all-order approximation change the results obtained in the CI approximation by only a few per cent (by 4.5\% for Si$^{2+}$ and by 8\% for Mg).

Since we use the numerically complete basis set and the saturated CI space, we take into account the valence-valence interactions almost exactly. The main source of uncertainty arises from the core-valence correlations. We conservatively estimate this uncertainty as the difference between the CI+MBPT and CI+All results presented in Table~\ref{tab:alph2}. The core contributions, $\alpha_2^c$, are very small for both Si$^{2+}$ and Mg (0.3\% for Si$^{2+}$ and less than 0.1\% for Mg). Even if we estimate (very conservatively) their uncertainties at the level of 10\%, their contribution to the uncertainty budget is negligible. Our final recommended values of $\alpha_2(3s^2~^1S_0)$  are 35.86(13) a.u. for Si$^{2+}$ and 814(3) a.u. for Mg. Note that in both cases our results are in excellent agreement with other most accurate results obtained by Mitroy in Ref.~\cite{Mit08} for Si$^{2+}$ and by Mitroy and Bromley in Ref.~\cite{MitBro03} for Mg.

\section{Conclusion}
In conclusion, we have developed a method for the precision calculation of electric quadrupole polarizabilities and applied it to evaluate the static E2 polarizabilities of the ground $3s^2\, ^1\!S_0$ state of the doubly-ionized Si and neutral Mg. Our recommended values are  $\alpha_2(^1\!S_0) = 35.86(13)$ a.u. for Si$^{2+}$ and 814(3) a.u. for Mg. To the best of our knowledge, these are the most accurate values of these quantities obtained
so far. They are in excellent agreement with the previous calculation of Porsev and Derevianko~\cite{PorDer06Z} for Mg and with the theoretical results obtained by CICP method in Refs.~\cite{Mit08,MitBro03}.

The method developed in this work can be used to calculate quadrupole
polarizabilities of the low-lying states of any divalent or trivalent system. We hope that the present work will stimulate experimental studies of quadrupole polarizabilities of divalent systems  using the resonant excitation Stark ionization spectroscopy and other methods for benchmark test of theory and experiment.

\section{ACKNOWLEDGEMENTS}
This work was supported in part by US NSF Grants No.\ PHY-1068699 and No.\ PHY-0758088. The work of MGK was supported in part by RFBR grant No.\
11-02-00943.


%

\end{document}